# Adjustable quantum interference oscillations in Sb-doped Bi$_2$Se$_3$ topological insulator nanoribbons

Hong-Seok Kim[1, ∥], Tae-Ha Hwang[1, ∥], Nam-Hee Kim[1], Yasen Hou[2], Dong Yu[2], H.-S. Sim[3], Yong-Joo Doh[1*]

[1]Department of Physics and Photon Science, School of Physics and Chemistry, Gwangju Institute of Science and Technology (GIST), Gwangju 61005, Korea

[2]Department of Physics, University of California, Davis, CA 95616, USA

[3]Department of Physics, Korea Advanced Institute of Science and Technology, Daejeon 34141, Korea

**Corresponding Authors**
*E-mail: yjdoh@gist.ac.kr

**Author contributions**
∥ These authors contributed equally to this work.

**Abstract**

Topological insulator (TI) nanoribbons (NRs) provide a unique platform for investigating quantum interference oscillations combined with topological surface states. One-dimensional subbands formed along the perimeter of a TI NR can be modulated by an axial magnetic field, exhibiting Aharonov–Bohm (AB) and Altshuler–Aronov–Spivak (AAS) oscillations of magnetoconductance (MC). Using Sb-doped $Bi_2Se_3$ TI NRs, we found that the relative amplitudes of the two quantum oscillations can be tuned by varying the channel length, exhibiting crossover from quasi-ballistic to diffusive transport regimes. The AB and AAS oscillations were discernible even for a 70-μm-long channel, while only the AB oscillations were observed for a short channel. Analyses based on ensemble-averaged fast Fourier transform of MC curves revealed exponential temperature dependences of the AB and AAS oscillations, from which the circumferential phase-coherence length and thermal length were obtained. Our observations indicate that the channel length in a TI NR can be a useful control knob for tailored quantum interference oscillations, especially for developing topological hybrid quantum devices.

The quantum interference effects of a charged particle's wave function provide the cornerstone for mesoscopic physics, as they affect the development of quantum information devices. The Aharonov–Bohm (AB) effect[1] is a central example. In the effect, a charged particle gains a gauge-invariant phase when it encircles magnetic flux, showing an interference signal known as AB oscillations (ABOs). ABOs appear in electrical conductance in various quasi-ballistic ring structures fabricated using normal metals[2], quantum wells[3], and graphene[4]. The conductance is modulated periodically, as a function of magnetic flux $\Phi$ through a loop, with a period of the flux quantum, i.e., $\Phi_0 = h/e$, where $h$ is the Planck constant, and $e$ is the elementary charge. On the other hand, in a weakly disordered ring[5] or a hollow metallic cylinder[6,7] including carbon nanotubes[8] and core/shell nanowires[9], the $h/2e$–periodic oscillations, which are known as Altshuler–Aronov–Spivak (AAS) oscillations[10], dominate $h/e$ oscillations. The AAS oscillations are caused by the interference between a pair of time-reversed paths enclosing the magnetic flux[7], which are formed by a sequence of elastic disorder scattering.

Topological insulator (TI) nanoribbons (NRs) provide a novel platform for investigating mesoscopic quantum interference effects combined with a nontrivial topological order[11,12]. Because three-dimensional (3D) TIs are bulk insulators with gapless spin-textured surface states[13], TI NRs can be regarded as hollow metallic cylinders, whose surface states are topologically protected from time-reversal invariant perturbations[13]. Considering quantum confinement along the perimeter of TI NRs, the surface states form discrete one-dimensional (1D) subbands exhibiting flux-dependent dispersion[14], which comprises the AB phase of $2\pi\Phi/\Phi_0$ owing to the axial magnetic field ($B_{axial}$) and the Berry phase of $\pi$ caused by the spin–momentum locking effect in TIs[11,12]. When the axial flux through the core of a TI NR becomes an odd multiple of the half-flux quantum, the AB phase shift cancels the Berry phase and hence restores a topologically protected zero-gap 1D mode[11, 12], which can host the Majorana bound states[15]. Resultantly the density of states of the surface states is modulated by $B_{\text{axial}}$ with a period of $\Phi_0$, resulting in magnetoconductance (MC) oscillations in TI NRs, whereas the MC maxima occur at the integer (0-ABO) or half-integer (π-ABO) multiples of $\Phi_0$, depending on the location of the Fermi level[11, 12].

Thus far, topological ABOs have been observed in various NRs made of $Bi_2Se_3$[14, 16, 17], Sb-doped $Bi_2Se_3$[18], $Bi_2Te_3$[19, 20, 21], and 3D Dirac semimetals[22, 23]. However, several unresolved issues remain. First, $h/2e$-periodic MC oscillations are concomitantly observed with ABOs in many cases[14, 16, 19, 23]; however, their absence has also been reported[18, 21, 22] without a clear explanation. Furthermore, for the $h/2e$-periodic MC oscillations, two conflicting explanations have been suggested: the weak antilocalization (WAL) effect along the perimeter of the TI NR in the diffusive regime[14, 16, 19] or the second harmonic of ABOs occurring in the quasi-ballistic regime[17, 20]. In addition, the temperature dependence of ABOs follows the diffusive[16, 19] or quasi-ballistic[14, 17, 20] transport behavior for the same kind of NRs with similar geometric dimensions. Finally, the crossover between AB and AAS oscillations in TI NRs is expected based on theory[11, 24]; however, its experimental verification has not been performed yet.

Herein, we present an extensive experimental study of AB and AAS oscillations obtained from Sb-doped $Bi_2Se_3$ TI NRs with various channel lengths ($L_{\text{ch}}$) on the same NR. Their axial MC oscillations were measured as a function of gate voltage ($V_g$) and temperature ($T$), whereas their oscillation amplitudes were analyzed using the ensemble-averaged fast Fourier transform (FFT) method to avoid sample-specific features. We observed the crossover behavior between the AB and AAS oscillations and discovered that their respective oscillation amplitudes were adjustable depending on the channel length of the TI NR. The AAS oscillations were absent for channel lengths shorter than the perimeter length ($L_p$), and their length dependence was consistent with theoretical expectations based on WAL corrections[11] along the perimeter of TI NR. The ABOs were discernible even for the 70 µm-long channel; furthermore, the $V_g$-dependent alternations of 0- and π-ABOs, a characteristic feature of topological ABOs[20], were evident in all segments with $L_{ch}$ = 1–70 µm. Our observations suggest that the $L_{ch}$ of the TI NR can be a control knob to alter the quantum electronic transport from quasi-ballistic to diffusive regimes, which would be advantageous for investigating quantum interference effects associated with topological surface states.

## Results and discussion

**Channel length dependence of AB and AAS oscillations.** Figure 1a shows the scanning electron microscopy (SEM) image of the Sb-doped $Bi_2Se_3$ TI NR device (**D2**) with various channel lengths fabricated on the same NR. The geometric dimensions of the TI NR are provided in Supporting Table 1. When an external magnetic field is applied along the NR axis, the axial MC, $G(B_{axial})$ curve shows oscillatory behavior superposed on a parabolic background (see Supporting Fig. 4). After subtracting the smooth background signal from the MC data, we obtained the periodic oscillations of MC variation, $\delta G(B_{axial})$, as shown in Fig. 1c. The oscillation period is obtained to be $\Delta B_{axial}$ = 0.075 T, which corresponds to Φ = 1.05Φ$_0$ considering a 5 nm-thick native oxide layer[17, 21] formed on the surface of the TI NR (see Supporting Fig. 2). In addition, two $\delta G(B_{axial})$ curves obtained at $V_g$ = -2.5 and -2.8 V show 0- and π-ABOs, respectively. Figure 1d displays the color plots of $\delta G(B_{axial})$ curves as a function of $V_g$ for different $L_{ch}$ values on the same TI NR. For $L_{ch}$ = 1 µm, the ABO phase alternates between 0 and π with varying $V_g$, exhibiting a checkerboard-like pattern with sharp boundaries. Similar rectangular patterns were obtained from different device (**D1**) with $L_{ch}$ = 1 µm over a wide range of $V_g$ (see Supporting Fig. 5), revealing the out-of-phase relationship between two

$\delta G(V_g)$ curves for $\Phi = 0$ and $0.5\Phi_0$ as supporting evidence for topological ABOs[11, 20] in the TI NRs. The FFT analysis for a 1-µm-long segment in **D2**, as depicted in Fig. 1e, shows a single peak corresponding to the $h/e$-periodic oscillations of $\delta G(B_{axial})$, indicating that the axial MC curves obtained from the short-channel devices were dominated by the ABOs only. Because the $\delta G(B_{axial})$ curve was extremely sensitive to $V_g$, the FFT spectrum in Fig. 1e represents the ensemble average of the total 51 FFT spectra obtained at different $V_g$ with a constant increment of $\Delta V_g = 100$ mV.

When the channel length was increased on the same TI NR, the overall amplitude of the $\delta G(B_{axial})$ curve decreased, and the shape of the checkerboard pattern deformed from rectangular to elliptical (see Fig. 1d). In particular, the positive $\delta G$ patterns elongated along the $V_g$ axis, whereas their widths reduced along the $B_{axial}$ axis. Such deformations in the $\delta G(B_{axial}, V_g)$ plots are attributed to the occurrence of $h/2e$-periodic MC oscillations in the long-channel devices, which are superposed on the topological ABO patterns in the TI NR. Contrary to the ABOs showing $V_g$-dependent phase alternation, the $h/2e$-periodic MC oscillations exhibited their conductance maxima at integer multiples of $\Phi_0/2$, irrespective of $V_g$. The $V_g$-independent behavior of the $h/2e$-periodic oscillations suggests that they were caused by the AAS effect rather than the second-harmonic AB effect.

The ensemble-averaged FFT results with different $L_{ch}$ values are displayed in Fig. 1e, exhibiting two peaks corresponding to the $h/e$- and $h/2e$-periodic oscillations. The peak heights of the FFT spectra are shown in Fig. 1f as a function of $L_{ch}$. The FFT peak height corresponding to the $h/e$-periodic oscillations decreases monotonically with increasing $L_{ch}$, whereas the $h/2e$-periodic oscillations exhibit a nonmonotonous channel-length dependence: nearly absent for $L_{ch} = 1$ µm, maximized at $L_{ch} = 2-3$ µm, and decreases gradually for $L_{ch} > 3$ µm. The absence of $h/2e$-periodic oscillations in the short-channel device is attributed to an incomplete formation of a pair of the time-reversed paths[11] along the perimeter of the TI NR for $L_{ch} < L_p$, where $L_p$ is 1.2 µm for **D2**. The solid line, which agrees well with the experimental data, is the best fit of the WAL correction[11] along the perimeter of the TI NR (see Methods), indicating that our observed $h/2e$-periodic oscillations for $L_{ch} > L_p$ were caused by AAS oscillations in the TI NR in the weak-disorder limit.

The decreasing amplitudes of the ABOs with increasing $L_{ch}$ can be explained by the effect of ensemble averaging of the uncorrelated ABOs in the TI NRs. Because the TI NRs are topologically analogous to hollow metallic cylinders[20], the long-channel device can be considered as a series array of metallic AB loops[5] exhibiting random phase shift of oscillations of $\delta G(B_{axial})$ at fixed $V_g$. Here, the number of loops, $N_{AB}$, is expressed as $N_{AB} = L_{ch}/L_\phi$, where $L_\phi$ is the phase coherence length[25]. We obtained $L_\phi = 465$ nm at $T = 3$ K from the WAL analysis of perpendicular MC in the TI NR (see Supporting Fig. 6). It is well known that the stochastic average of ABOs in a chain of $N_{AB}$ loops results in the suppression of the oscillation amplitude, which is proportional to $N_{AB}^{-3/2}$, considering the connecting leads between the loops[5, 25] ($N_{AB}^{-1/2}$ from average over the $N_{AB}$ oscillations with random phase shift, and additional $N_{AB}^{-1}$ from classical resistance along the length $L_{ch}$). Hence, we fitted a $L_{ch}^{-3/2}$ function to the ensemble-averaged FFT peak heights of the $h/e$-periodic oscillations (the dashed line in Fig. 1f), and it agrees well with the experimental data except $L_{ch} = 1$ µm.

More interestingly, the AB and AAS oscillations were also observed in an extremely long NR segment of $L_{ch} = 70$ µm, as shown in Fig. 2a. To the best of our knowledge, this is the longest TI NR exhibiting ABOs thus far. The $\delta G(B_{axial})$ curve in Fig. 2c reveals the coexistence of the AB and AAS oscillations, exhibiting the $V_g$-dependent alternation of 0- and π-ABOs as well. The color plot of $\delta G(B_{axial}, V_g)$ shows the checkerboard pattern of the ABOs overlaid with the AAS oscillations. The $h/2e$-periodic oscillations became insignificant at $B_{axial}$ fields greater than ~ 0.5 T, which was induced by the suppression of coherent backscattering due to time-reversal symmetry breaking[9, 14]. By contrast, the ABOs persisted up to higher $B_{axial}$ fields, as depicted in Fig. 2c, because the AB phase was due to the coherent forward scattering of the surface electrons on the TI NRs. It is noteworthy that the TI NR segment with $L_{ch}$ = 70 µm corresponds to a chain of AB loops with $N_{AB}$ ~ 151, indicating that the visibility of the ABOs in the TI NR is robust to the ensemble average compared with the metallic loops[5]. The robustness of the TI NR is attributable to the extremely thin topological-surface-state thickness (i.e., < 6 nm, as estimated from the full-width at half maximum of the FFT spectral peak[16]) and the uniform cross-sectional area of the TI NR used in this study (see Supporting Fig. 3). Moreover, the modulation of the density of states in the 1D subbands of the TI NR, which is due to the zero-gap 1D mode occurring at half-integer multiples of $\Phi_0$, is responsible for the robust ABOs. For the segment with $L_{ch}$ = 5 µm, similar features were observed, as shown in Fig. 2b, except the relative ratio between the AB and AAS oscillation amplitudes.

The ensemble-averaged FFT spectra with different $L_{ch}$ values, normalized by the sum of the two peak heights of $h/e$- and $h/2e$-periodic oscillations, are shown in Fig. 2d. As mentioned previously, the ABOs dominated over the 1-µm-long channel device with $N_{AB}$ ~ 2, which is characteristic of quasi-ballistic transport in TI NRs. By contrast, the quantum interference oscillations in the 70-µm-long channel with $N_{AB}$ ~ 151 were dominated by AAS oscillations with WAL corrections, indicating that the long-channel TI NRs exhibited diffusive quantum transport at low temperatures. For the 5-µm-long channel ($N_{AB}$ ~ 11), the FFT peak height of the AB oscillations was similar to that of the AAS oscillations, indicating a crossover between ballistic and diffusive quantum transport in the TI NR. Although temperature- or disorder-driven crossover between AB and AAS oscillations has been predicted theoretically[26,

[27], it has not been demonstrated experimentally yet. Our observations indicate that those quantum interference oscillations can be tuned by adjusting the channel length of the TI NR in the weak-disorder limit. Meanwhile, the ensemble-averaged FFT spectrum can be used to identify the electrical transport regime of the TI NRs at an arbitrary disorder strength. We expect the channel-length-dependent topological quantum interferometers of the TI NR to be useful for investigating various features of topological quantum devices combined with superconductivity[28, 29], ferromagnetism[30, 31], or nanomechanics[32].

**Temperature dependence of AB and AAS oscillations.** The color plots of $\delta G(B_{axial}, V_g)$ with different $L_{ch}$ values as a function of temperature are shown in Figs. 3a–c. The overall amplitude of $\delta G(B_{axial}, V_g)$ diminished owing to thermal fluctuations. Furthermore, the boundaries of the checkerboard patterns became highly irregular at higher temperatures for $L_{ch} = 2$ µm in **D2**, as shown in Fig. 3a. However, the other segments in **D6**, maintained their pattern shapes at high temperatures (see Figs. 3b–c), which will be discussed later. The ensemble-averaged FFT spectra obtained at different temperatures are displayed in Figs. 3d–f, revealing two peaks corresponding to the $h/e$- and $h/2e$-periodic oscillations. The FFT amplitudes for the $\delta G(B_{axial})$ curves with different $V_g$ were averaged over 21 (27) traces with the increment of $\Delta V_g = 100$ (50) mV for the 2-µm-long segment in **D2** (5- and 70-µm-long segments in **D6**).

Figures 3g–i show the heights of the two FFT peaks as a function of temperature. Linear $T$ dependences of the FFT peak heights in the semi-log plot indicate that the amplitudes of the AB and AAS oscillations in the TI NRs, regardless of $L_{ch}$, decay exponentially with temperature, resulting in $\delta G_i(T) \sim \exp(-b_i T)$, where $b_i$ is the damping parameter ($i$ signifies $h/e$ or $h/2e$ oscillations) of each quantum oscillation. Previous studies using the single-trace FFT method reported exponential temperature dependence[14, 17, 20, 21] or a $T^{-1/2}$ behavior[16, 19] for the amplitude of the ABOs in TI NRs. In this study, we observed the exponential $T$ dependence of both AB and AAS oscillations using the ensemble-averaged FFT method to avoid any confusion caused by universal conductance fluctuations[33] or $V_g$-dependent voltage fluctuations due to residual charges on the substrate[34].

The exponential temperature dependence of the ABOs amplitude, which was also observed in a ballistic AB ring[34, 35] comprising two-dimensional electron gas, is attributed to the thermal averaging effect of disorder-induced phase shifts near the Fermi energy[35, 36]. The thermal broadening effect can be described by $\exp(-b_{h/e}T) = \exp(-k_B T/E_c)$, where $k_B$ is the Boltzmann constant, $E_c = \hbar v_F/L_p$ the correlation energy, $\hbar$ the reduced Planck's constant, and $v_F$ the Fermi velocity[14, 35]. For $L_{ch} = 2$ μm in **D2**, we obtained $E_c = 165$ μeV using $L_p = 1.2$ μm and $v_F = 3 \times 10^5$ m/s (see Supporting Note 2). Subsequently, the AB damping parameter was calculated to be $b_{h/e,cal} = k_B/E_c = 0.52$ K$^{-1}$, which is close to the experimental value of $b_{h/e} = 0.56$ K$^{-1}$ in Fig. 3g. For other NR segments in **D6**, we obtained $E_c = 251$ μeV and $b_{h/e,cal} = 0.34$ K$^{-1}$ using $L_p = 0.79$ μm, which agreed well with the experimental values of $b_{h/e} = 0.35$ and $0.34$ K$^{-1}$ for $L_{ch} = 5$ and $70$ μm, respectively, as shown in Figs. 3h–i. The smaller $L_p$ for the segments in **D6** resulted in a larger $E_c$ than that obtained from **D2**. Based on the expression of $\exp(-b_{h/e}T) = \exp(-L_p/L_T(T))$, where $L_T$ means the disorder-induced thermal length, we obtained $L_T(T = 1$ K$) = 2.1$ and $2.3$ μm, corresponding to $1.8$ and $2.9$ $L_p$, for **D2** and **D6**, respectively; hence, the ABOs in **D6** were robust against thermal fluctuations and preserved the pattern shapes at high temperatures, as depicted in Figs. 3b–c.

Although the amplitude of the AAS oscillations exhibited a similar exponential behavior of $\exp(-b_{h/2e}T)$, as shown in Figs. 3g–i, its underlying mechanism differed from that of the $h/e$-periodic oscillations. Because the AAS oscillations are due to the quantum interference between a pair of time-reversed paths formed by a sequence of elastic disorder scattering, the phase shift for the same path is zero and the disorder effects are cancelled out;

therefore, the AAS oscillations are insensitive to the thermal broadening effect[35]. By contrast, thermal dephasing through inelastic electron–electron scattering results in a shorter phase coherence length[35]; thus, the expression $\exp(-b_{h/2e}T) = \exp(-2L_p/L_{\varphi,c}(T))$ holds, where $L_{\varphi,c}$, which is the phase coherence length along the circumference of the TI NR, was used instead of $L_\varphi$ estimated from the perpendicular MC data. Using the relation $L_{\varphi,c} = 2L_p/b_{h/2e}T$, the circumferential coherence length at $T = 1$ K is estimated to be $L_{\varphi,c}(T = 1\text{ K}) = 2.8, 4.2$, and $3.4$ µm for $L_{ch} = 2, 5$, and 70 µm, respectively, from the experimental values of $b_{h/2e}$ in Figs. 3g–i. It is noteworthy that our observed $L_{\varphi,c}(T = 1\text{ K})$ is two to five times longer than $L_p$, which is sufficient to assure phase-coherent quantum transport along the circumference of the TI NRs. Moreover, our maximum $L_{\varphi,c}$ value is several times larger than those reported previously[17, 20], indicating that the TI NRs in this study are in the weak-disorder limit.

## Conclusions

In summary, we studied the channel length dependence of AB and AAS oscillations in Sb-doped $Bi_2Se_3$ TI NRs. Our observations demonstrate that the relative amplitudes of the AB and AAS oscillations can be adjusted by the channel length in comparison with the perimeter length of TI NR. Two quantum inteference oscillations are clearly observed even in a 70-µm-long channel device in diffusive regime, while the AAS oscillations are absent in short-channel devices in quasi-ballistic transport regime. The thermal broadening induced by elastic disorder scattering and thermal decoherence effect due to, e.g., inelastic electron-electron interactions are responsible for the exponential temperature dependence of the AB and AAS oscillations, respectively. Our observations suggest that the channel length in TI NR can be a useful tool for tailoring quantum interference effects combined with topological surface states.

## Methods

**Device fabrication.** Sb-doped $Bi_2Se_3$ NRs were synthesized via chemical vapor deposition method in a horizontal tube furnace. Detailed information is available elsewhere[37]. Energy-dispersive X-ray spectroscopy of the NRs revealed the atomic percentages of Bi, Sb, and Se of approximately 36.0%, 5.5%, and 58.5%, respectively (see Supporting Fig. 1). After the growth was completed, individual $(Bi_{1-x}Sb_x)_2Se_3$ NRs were mechanically transferred onto a highly *n*-doped Si substrate covered with a 300-nm-thick $SiO_2$ layer. The Si substrate was used as a back gate electrode. Source and drain electrodes were defined using standard electron-beam lithography followed by the electron-beam evaporation of Ti (10 nm)/Au (200 nm). Prior to the metal deposition, the electron-beam resist residue and the native oxide layer on the surface of NR were removed using oxygen plasma treatment and by dipping into a 6:1 buffered oxide etch for 7 s.

**Measurements.** All electrical transport measurements were performed using a conventional lock-in technique in a four-probe configuration. We used a closed-cycle $^4$He cryostat (Seongwoo Instruments Inc.) and $^3$He refrigerator system (Cryogenic, Ltd.), which had base temperatures of 2.4 and 0.3 K, respectively.

**Weak antilocalization correction along the perimeter of TI NR.** The WAL correction using the boundary conditions of the cylindrical geometry is as follows:[11]

$$\delta G = \frac{L_p}{L_{ch}} \frac{e^2}{\pi h} \left[ \log \frac{L_{ch}}{\xi} + \sum_{n=1}^{\infty} \cos \frac{4\pi n \Phi}{\Phi_0} \log \left(1 - e^{-\pi n L_p / L_{ch}}\right) \right],$$

where $L_p$ is the perimeter length of the NR, $L_{ch}$ the channel length, and $\xi$ the correlation length of the disorder potential. The line of best fit in Fig. 1f was obtained using $\xi = 850$ nm as a fitting parameter and $L_p$ = 1.2 µm (for **D2**) considering a 5-nm-thick oxide layer[17, 21] on the surface (see Supporting Fig. 2).

# Figure Captions

**Figure 1 | Channel length dependence of *h/e*- and *h/2e*-periodic oscillations.** (**a**) SEM image of Sb-doped $Bi_2Se_3$ NR device (D2) with various $L_{ch}$ in the same NR. Channel lengths between the electrodes were $L_{ch}$ = 1, 2, 3, and 4 µm for electrode pairs 1-2, 3-4, 5-6, and 6-7, respectively. Those pairs were used to measure voltage differences while a bias current was flowing through the entire NR. (**b**) Tilted-view SEM image of NR segment between 3-4 electrodes. (**c**) Conductance variation, $\delta G$, *vs.* axial magnetic field, $B_{axial}$ curves obtained from NR segment (1-2) with $L_{ch}$ = 1 µm for two different gate voltages, $V_g$ at $T$ = 2.4 K. Magnetic flux is denoted by $\Phi/\Phi_0$. (**d**) Color plot of $\delta G$ as a function of $B_{axial}$ and $V_g$ for different $L_{ch}$. (**e**) Ensemble-averaged FFT amplitudes of $\delta G(B_{axial})$ curves for different $L_{ch}$. (**f**) FFT peak heights corresponding to *h/e*- and *h/2e*-periodic oscillations as a function of $L_{ch}$. The solid line is from the theoretical fit to the WAL correction (see Methods), whereas the dashed one from $L_{ch}^{-3/2}$ dependence (see text).

**Figure 2 | *h/e*- and *h/2e*-periodic oscillations in long-channel devices.** (**a**) SEM image of $(Bi_{0.89}Sb_{0.11})_2Se_3$ NR device (D6) with channel lengths of $L_{ch}$ = 5 and 70 µm. Bias current was applied between a pair of electrodes numbered 1 and 5, whereas voltage difference was measured between electrodes 3 and 4 (2 and 3) for $L_{ch}$ = 70 (5) µm. $\delta G(B_{axial})$ curves at different $V_g$ and color plot of $\delta G(B_{axial}, V_g)$ at 2.6 K for (**b**) $L_{ch}$ = 5 µm and (**c**) $L_{ch}$ = 70 µm. $\delta G(B_{axial})$ curves were offset vertically for clarity. (**d**) Normalized ratio plot of ensemble-averaged FFT spectra with various $L_{ch}$. *h/e* (*h/2e*) oscillations are indicated by red (blue) arrows.

**Figure 3 | Temperature dependence of FFT for *h/e* and *h/2e* period oscillations.** Color plot of $\delta G(B_{axial}, V_g)$ at different temperatures for (**a**) $L_{ch}$ = 2 µm, (**b**) 5 µm, and (**c**) 70 µm. Ensemble-averaged FFT amplitudes of $\delta G(B_{axial})$ curves at different temperatures for (**d**) $L_{ch}$ = 2 µm, (**e**) 5 µm, and (**f**) 70 µm. FFT curves were averaged for all measured $V_g$. Temperature dependence of FFT peak heights of *h/e* and *h/2e* oscillations for (**g**) $L_{ch}$ = 2 µm, (**h**) 5 µm, and (**i**) 70

µm. Solid lines are best fit results with damping parameters $b_i$ ($i = h/e$, $h/2e$) (see text).

# References

1. Aharonov Y, Bohm D. Significance of Electromagnetic Potentials in the Quantum Theory. *Physical Review* **115**, 485-491 (1959).

2. Webb RA, Washburn S, Umbach CP, Laibowitz RB. Observation of *h/e* Aharonov-Bohm Oscillations in Normal-Metal Rings. *Physical Review Letters* **54**, 2696-2699 (1985).

3. Datta S, *et al.* Novel Interference Effects between Parallel Quantum Wells. *Physical Review Letters* **55**, 2344-2347 (1985).

4. Russo S, *et al.* Observation of Aharonov-Bohm conductance oscillations in a graphene ring. *Physical Review B* **77**, 085413 (2008).

5. Umbach CP, Van Haesendonck C, Laibowitz RB, Washburn S, Webb RA. Direct observation of ensemble averaging of the Aharonov-Bohm effect in normal-metal loops. *Physical Review Letters* **56**, 386-389 (1986).

6. Sharvin DY, Sharvin YV. Magnetic-flux quantization in a cylindrical film of a normal metal. *JETP Letters* **34**, 272-275 (1981).

7. Aronov AG, Sharvin YV. Magnetic flux effects in disordered conductors. *Reviews of Modern Physics* **59**, 755-779 (1987).

8. Bachtold A, *et al.* Aharonov-Bohm oscillations in carbon nanotubes. *Nature* **397**, 673-675 (1999).

9. Jung M, *et al.* Quantum Interference in Radial Heterostructure Nanowires. *Nano Letters* **8**, 3189-3193 (2008).

10. Al'tshuler B, Aronov A, Spivak B. The Aharonov-Bohm effect in disordered conductors. *JETP Letters* **33**, 94 (1981).

11. Bardarson JH, Brouwer PW, Moore JE. Aharonov-Bohm oscillations in disordered topological insulator nanowires. *Physical Review Letters* **105**, 156803 (2010).

12. Zhang Y, Vishwanath A. Anomalous Aharonov-Bohm conductance oscillations from topological insulator surface states. *Physical Review Letters* **105**, 206601 (2010).

13. Hasan MZ, Kane CL. Colloquium: Topological insulators. *Reviews of Modern Physics* **82**, 3045-3067 (2010).

14. Hong SS, Zhang Y, Cha JJ, Qi XL, Cui Y. One-dimensional helical transport in topological insulator nanowire interferometers. *Nano Letters* **14**, 2815-2821 (2014).

15. Cook A, Franz M. Majorana fermions in a topological-insulator nanowire proximity-coupled to an s-wave superconductor. *Physical Review B* **84**, 201105 (2011).

16. Peng H, *et al.* Aharonov-Bohm interference in topological insulator nanoribbons. *Nature Materials* **9**, 225-229 (2010).

17. Dufouleur J, *et al.* Quasiballistic Transport of Dirac Fermions in a $Bi_2Se_3$ Nanowire. *Physical Review Letters* **110**, 186806 (2013).

18. Cho S, *et al.* Aharonov-Bohm oscillations in a quasi-ballistic three-dimensional topological insulator nanowire. *Nature Communications* **6**, 8634 (2015).

19. Xiu F, *et al.* Manipulating surface states in topological insulator nanoribbons. *Nature Nanotechnology* **6**, 216-221 (2011).

20. Jauregui LA, Pettes MT, Rokhinson LP, Shi L, Chen YP. Magnetic field-induced helical mode and topological transitions in a topological insulator nanoribbon. *Nature Nanotechnology* **11**, 345-351 (2016).

21. Kim H-S, Shin HS, Lee JS, Ahn CW, Song JY, Doh Y-J. Quantum electrical transport properties of topological insulator $Bi_2Te_3$ nanowires. *Current Applied Physics* **16**, 51-56 (2016).

22. Kim J, *et al.* Quantum Electronic Transport of Topological Surface States in $β$-$Ag_2Se$ Nanowire. *ACS Nano* **10**, 3936-3943 (2016).

23. Wang L-X, Li C-Z, Yu D-P, Liao Z-M. Aharonov–Bohm oscillations in Dirac semimetal $Cd_3As_2$ nanowires. *Nature Communications* **7**, 10769 (2016).

24. Sacksteder IV VE, Wu Q. Quantum interference effects in topological nanowires in a longitudinal magnetic field. *Physical Review B* **94**, 205424 (2016).

25. Washburn S, Webb RA. Aharonov-Bohm effect in normal metal quantum coherence and transport. *Advances in Physics* **35**, 375-422 (1986).

26. Stone AD, Imry Y. Periodicity of the Aharonov-Bohm effect in normal-metal rings. *Physical Review Letters* **56**, 189 (1986).

27. Bardarson JH, Moore JE. Quantum interference and Aharonov-Bohm oscillations in topological insulators. *Reports on Progress in Physics* **76**, 056501 (2013).

28. Kayyalha M, *et al.* Anomalous Low-Temperature Enhancement of Supercurrent in Topological-Insulator Nanoribbon Josephson Junctions: Evidence for Low-Energy Andreev Bound States. *Physical Review Letters* **122**, 047003 (2019).

29. Kim N-H, Kim H-S, Hou Y, Yu D, Doh Y-J. Superconducting quantum interference devices made of Sb-doped $Bi_2Se_3$ topological insulator nanoribbons. *Current Applied Physics* **20**, 680 (2020).

30. Mellnik AR, *et al.* Spin-transfer torque generated by a topological insulator. *Nature* **511**, 449 (2014).

31. Hwang T-H, Kim H-S, Kim H, Kim JS, Doh Y-J. Electrical detection of spin-polarized current in topological insulator $Bi_{1.5}Sb_{0.5}Te_{1.7}Se_{1.3}$. *Current Applied Physics* **19**, 917 (2019).

32. Kim M, *et al.* Nanomechanical characterization of quantum interference in a topological insulator nanowire. *Nature Communications* **10**, 4522 (2019).

33. Lee PA, Stone AD. Universal Conductance Fluctuations in Metals. *Physical Review Letters* **55**, 1622 (1985).

34. Lin K-T, Lin Y, Chi CC, Chen JC, Ueda T, Komiyama S. Temperature- and current-dependent dephasing in an Aharonov-Bohm ring. *Physical Review B* **81**, 035312 (2010).

35. Hansen AE, Kristensen A, Pedersen S, Sørensen CB, Lindelof PE. Mesoscopic decoherence in Aharonov-Bohm rings. *Physical Review B* **64**, 045327 (2001).

36. Kobayashi K, Aikawa H, Katsumoto S, Iye Y. Probe-Configuration-Dependent Decoherence in an Aharonov–Bohm Ring. *Journal of the Physical Society of Japan* **71**, 2094 (2002).

37. Hou Y, *et al.* Millimetre-long transport of photogenerated carriers in topological insulators. *Nature Communications* **10**, 5723 (2019).

## Acknowledgements

This study was supported by the NRF of Korea through the Basic Science Research Program (2018R1A3B1052827) and the SRC Center for Quantum Coherence in Condensed Matter (2016R1A5A1008184), and by the "GIST-Caltech Research Collaboration" grant funded by GIST. The work at UC Davis was supported by the U.S. National Science Foundation (Grant DMR-1838532). We are grateful to A. Morpurgo, K. Kang, J. H. Bardarson, and L. Rokhinson for their useful discussions.

## Author contributions

H.-S.K. and T.-H.H. contributed equally to this study.

## Competing interests

The authors declare no competing interests.

## Additional information

Correspondence and requests for materials should be addressed to Y.-J.D.

**Figure 1**

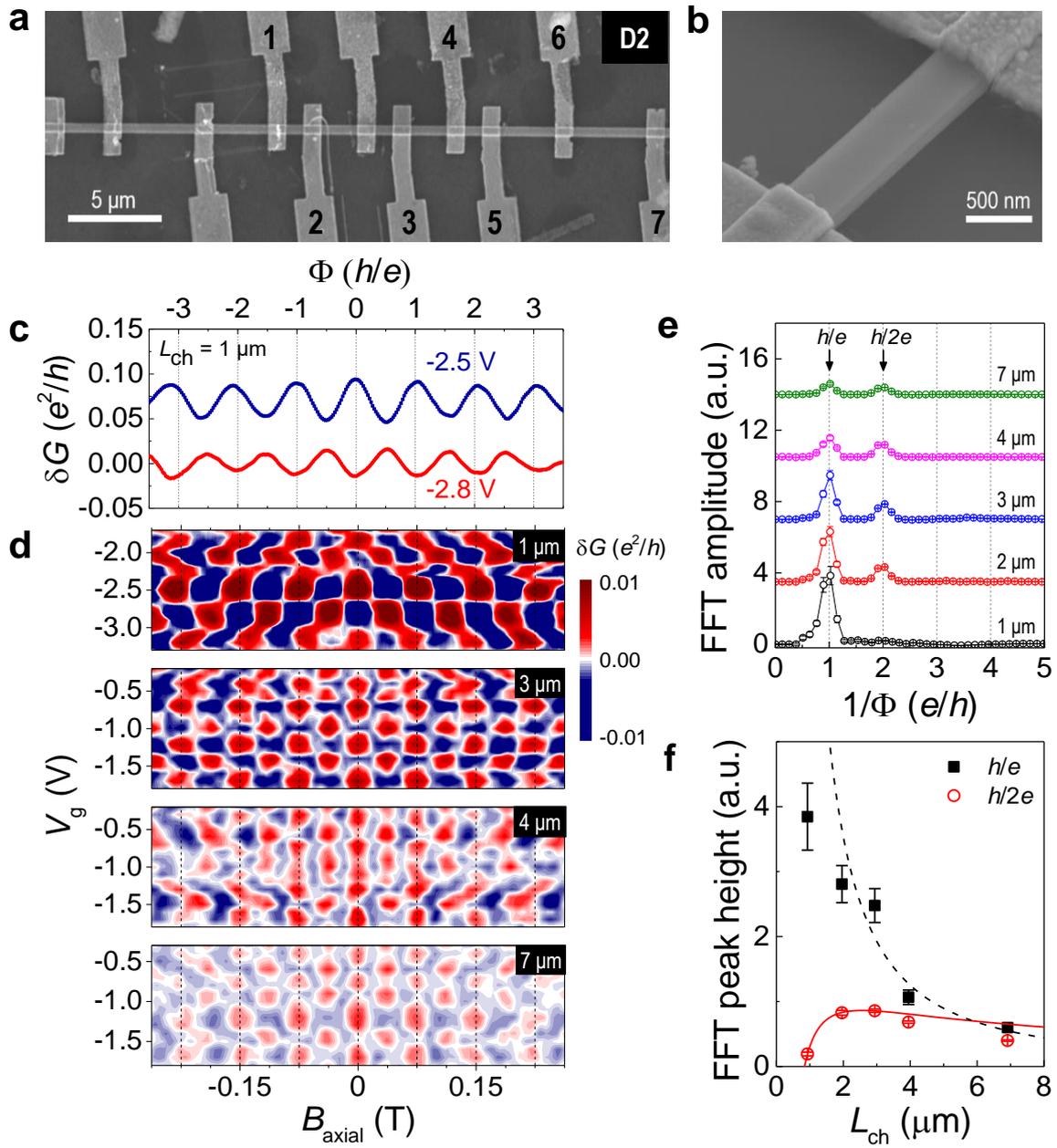

**Figure 2**

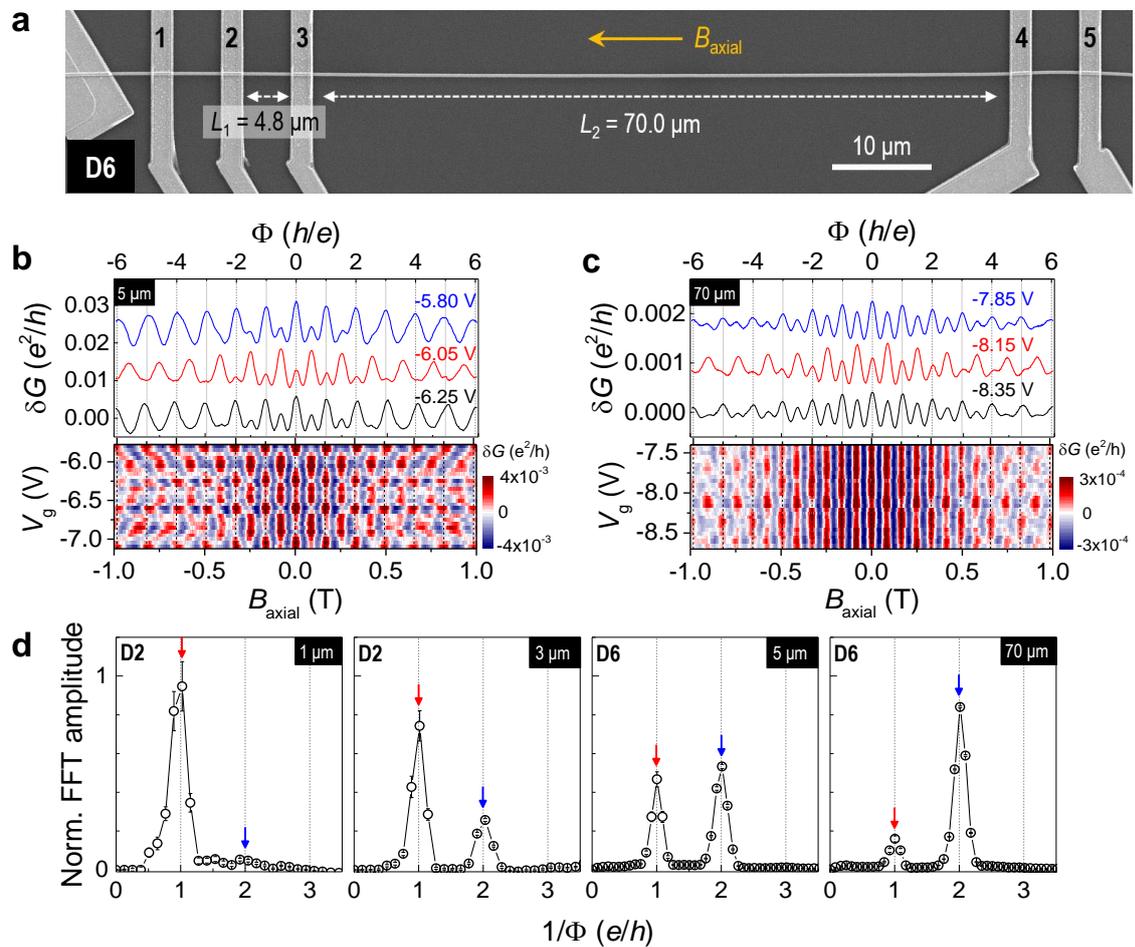

**Figure 3**

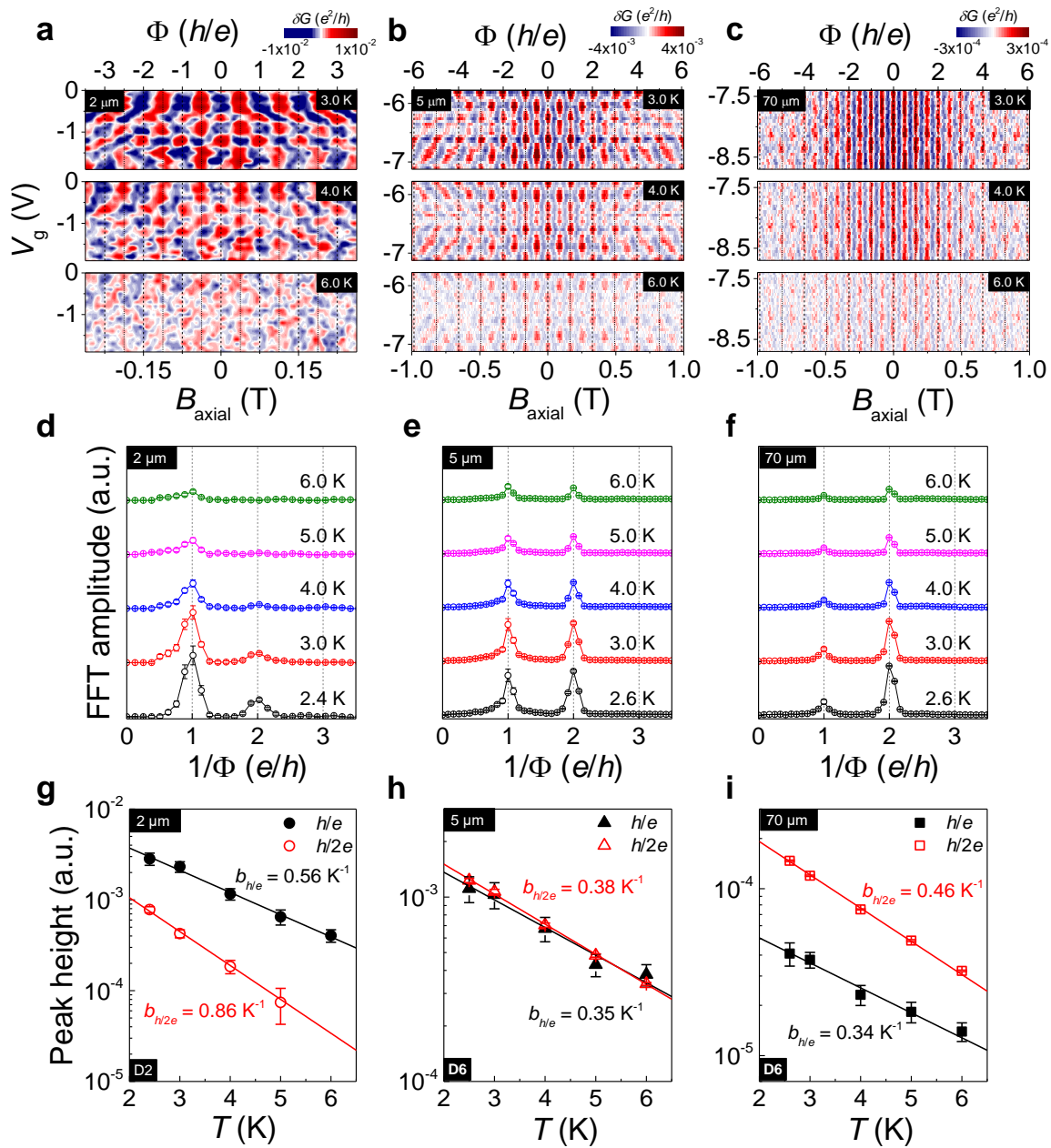

SUPPORTING INFORMATION FOR:

# Adjustable quantum interference oscillations in Sb-doped $Bi_2Se_3$ topological insulator nanoribbons

Hong-Seok Kim[1,∥], Tae-Ha Hwang[1,∥], Nam-Hee Kim[1], Yasen Hou[2], Dong Yu[2], H.-S. Sim[3], Yong-Joo Doh[1,*]

[1]Department of Physics and Photon Science, School of Physics and Chemistry, Gwangju Institute of Science and Technology (GIST), Gwangju 61005, Korea

[2]Department of Physics, University of California, Davis, CA 95616, USA

[3]Department of Physics, Korea Advanced Institute of Science and Technology, Daejeon 34141, Korea

**Corresponding Authors**
[*]E-mail: yjdoh@gist.ac.kr

**Author contributions**
[∥] These authors contributed equally to this work.

1. Supporting Table 1. Dimensional parameters of Sb-doped $Bi_2Se_3$ NR devices
2. Supporting Figure 1. EDS data of Sb-doped $Bi_2Se_3$ NRs
3. Supporting Figure 2. TEM image of Sb-doped $Bi_2Se_3$ NR
4. Supporting Figure 3. SEM and AFM image of Sb-doped $Bi_2Se_3$ NR devices
5. Supporting Figure 4. $G(B_{axial})$ raw data for D2 with $L_{ch} = 1$ μm
6. Supporting Figure 5. Gate-voltage dependence of ABOs for D1 with $L_{ch} = 1$ μm
7. Supporting Figure 6. Angle and temperature dependence of WAL
8. Supporting Figure 7. SdH oscillations
9. Supporting Note 1. Weak antilocalization (WAL) effect
10. Supporting Note 2. Shubnikov-de Haas (SdH) oscillations

# Supporting Table

| Sample | Width (nm) | Thickness (nm) | Total length (µm) |
| --- | --- | --- | --- |
| D1 | 607 | 90 | 35 |
| D2 | 468 (top), 520 (bottom) | 130 | 39 |
| D3 | 486 | 179 | 45 |
| D4 | 400 | 80 | 27 |
| D5 | 872 | 220 | 130 |
| D6 | 325 | 90 | 150 |

**Supporting Table 1.** Geometric parameters of Sb-doped $Bi_2Se_3$ NR devices.

**Supporting Figures:**

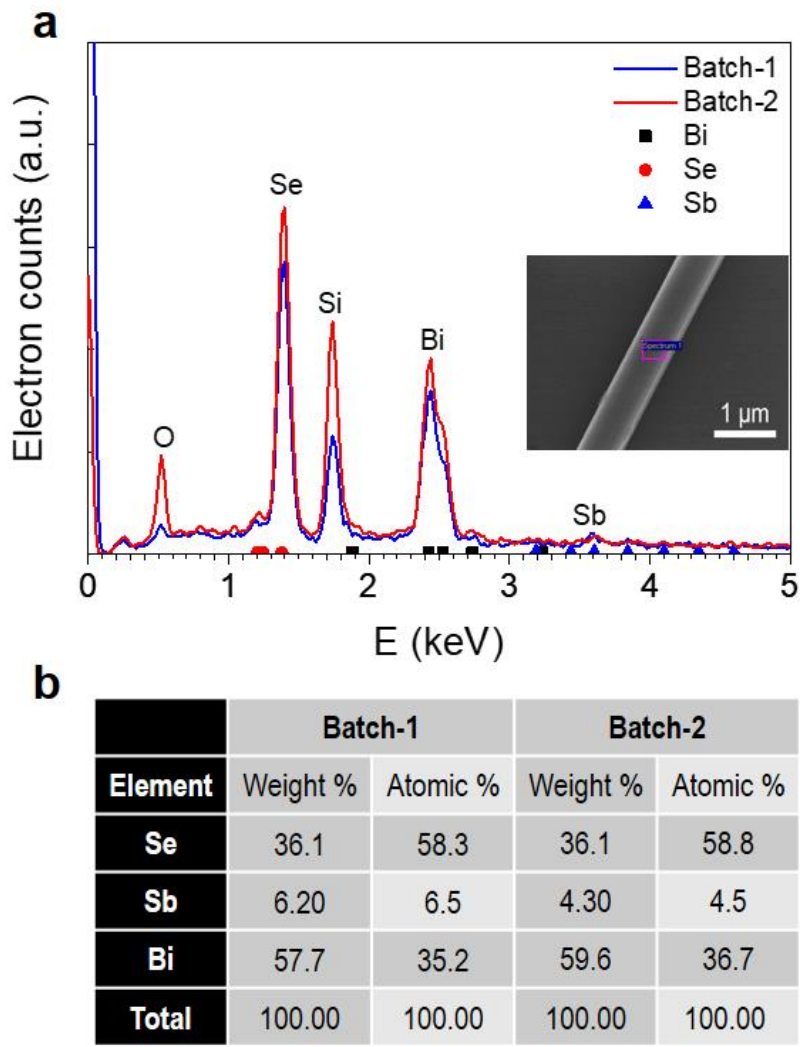

**Supporting Figure 1.** (a) Energy-dispersive X-ray spectroscopy (EDS) of Sb-doped $Bi_2Se_3$ NRs. Batch-1 (batch-2) was used for devices D1−D5 (D6). Inset: SEM image of TI NR. (b) Averaged values of weight percentage and atomic percentage of the elements of TI NRs from batch-1 and batch-2, respectively.

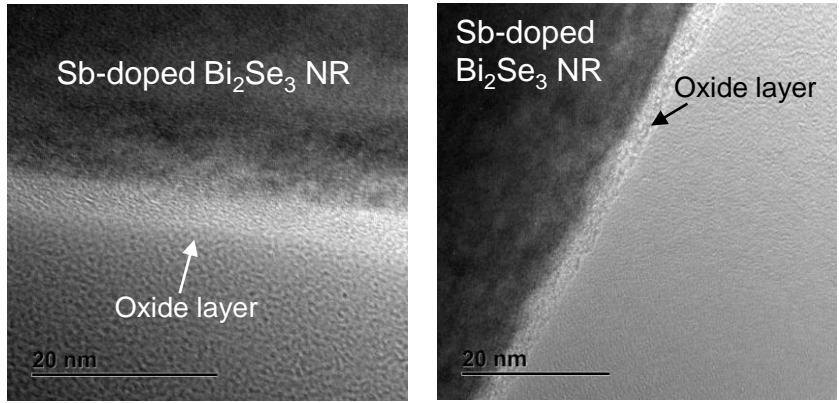

**Supporting Figure 2.** Transmission electron microscopy (TEM) image of TI NR.

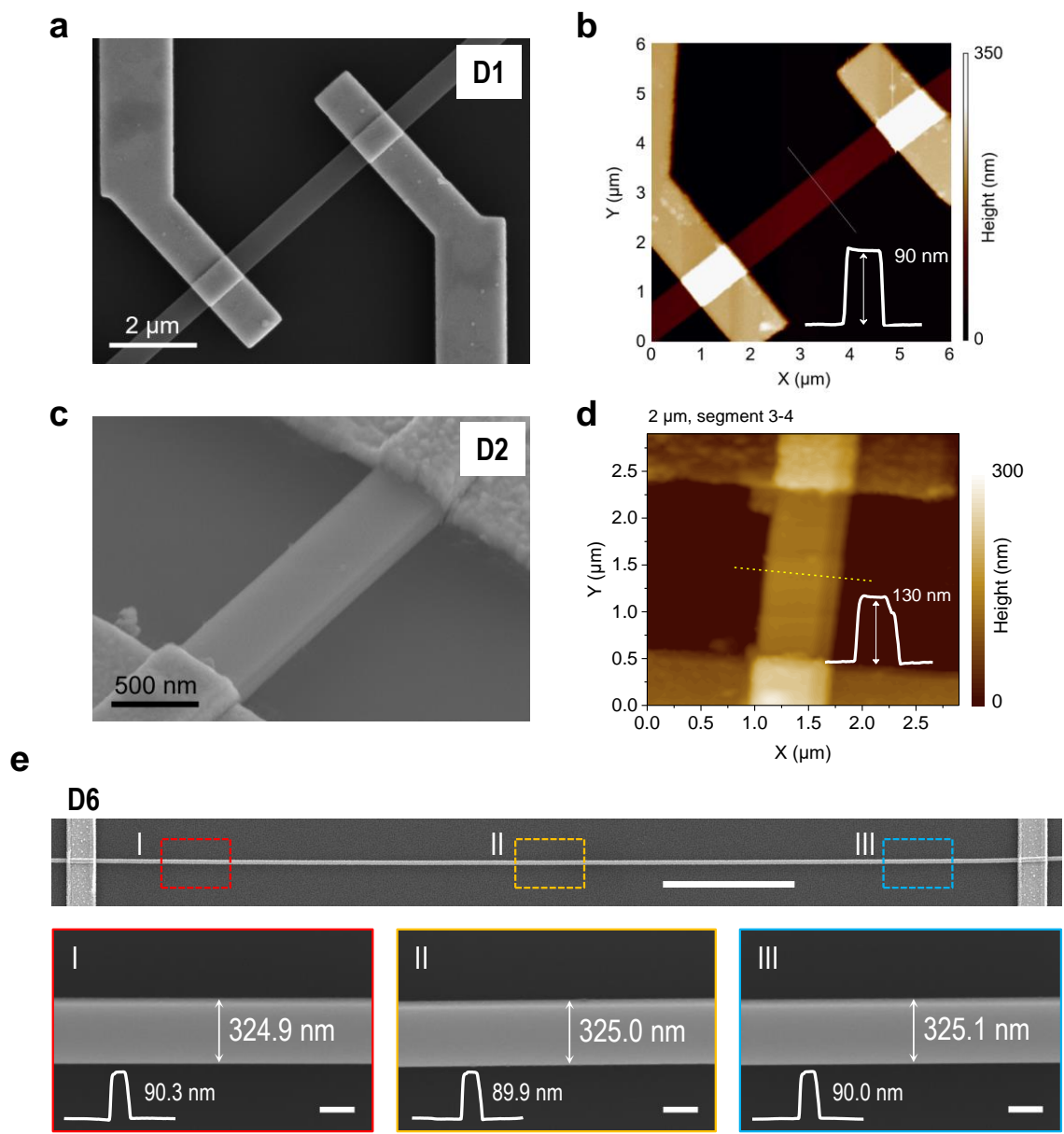

**Supporting Figure 3.** (a) SEM and (b) atomic force microscope (AFM) image of the device D1. Thickness of TI NR is obtained to be 90 nm. (c) SEM and (d) AFM image of the device D2. The thickness of the TI NR was measured to be 130 nm along the dashed line. (e) SEM image of the device D6 with a channel length of 70 μm. Scale bar is 10 μm. Bottom: Magnified views of the rectangular areas corresponding to the region-I, -II, and –III, respectively. Scale bars are 200 nm. Insets: AFM line profiles of the TI NR in each region.

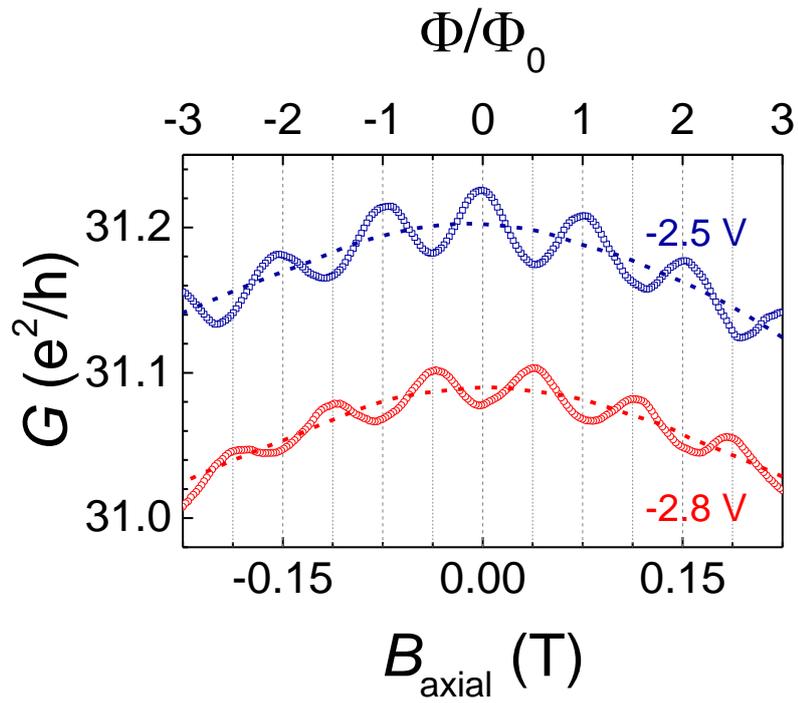

**Supporting Figure 4.** Conductance, *G*, *vs*. axial magnetic field, $B_{axial}$, curves of device D2 measured at *T* = 2.4 K for two different gate voltages, $V_g$ = -2.5 and -2.8 V. The *x*-axis on top represents the magnetic flux normalized by a flux quantum, $\Phi_0 = h/e$. Dashed line indicates a parabolic background for each $G(B_{axial})$ curve.

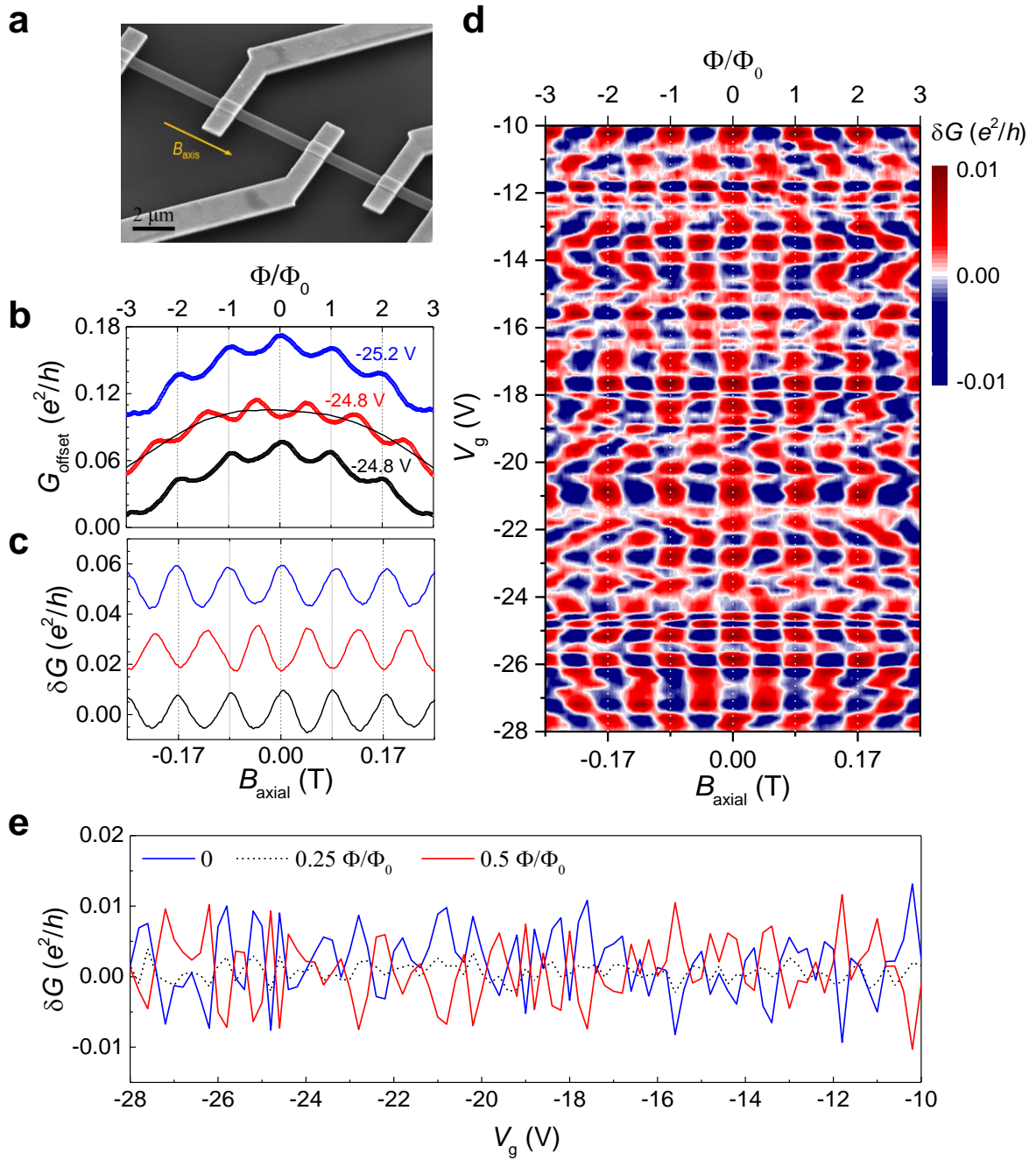

**Supporting Figure 5.** Gate-voltage dependence of ABOs. (a) SEM image of the device D1. (b) $G$ vs. $B_{axial}$ curves for the segment with $L_{ch}$ = 1 μm, measured at $T$ = 3.0 K for different gate voltages, $V_g$. Top $x$-axis represents the normalized magnetic flux, $\Phi/\Phi_0$. Black solid line indicates a parabolic background. (c) Differential conductance, $\delta G(B_{axial})$, curves with different $V_g$. Those curves are offset vertically for clarity. (d) Color plot of $\delta G$ as a function of $B_{axial}$ and $V_g$. (e) Line plot of $\delta G(V_g)$ curves for $\Phi/\Phi_0$ = 0, 0.25, and 0.5, respectively.

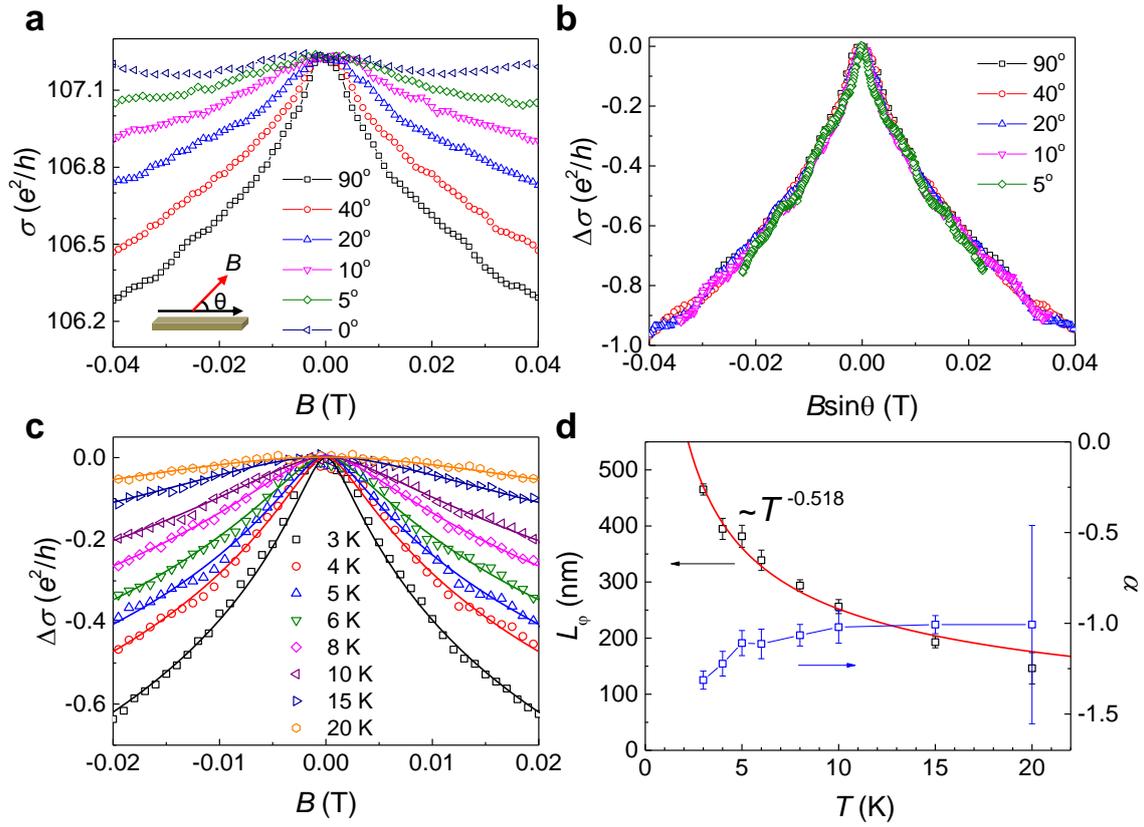

**Supporting Figure 6.** (a) Angle-dependent magnetoconductivity (MC) of the Sb-doped $Bi_2Se_3$ NR device, D1, with $L_{ch}$ = 5 μm at $T$ = 3 K. Inset: schematic of the measurement configuration. (b) MC variation, $\Delta\sigma(B) = \sigma(B) - \sigma(0)$, as a function of the perpendicular magnetic field component, $B\sin\theta$. The MC variation curves are merged into a single one at low fields. (c) MC variation curves measured at different temperatures with $\theta$ = 90° at $V_g$ = 0 V. The black solid line is a fit to the weak anti-localization (WAL) theory (see Supporting Note 1). (d) Temperature dependence of $L_\phi$ and $\alpha$, obtained from the WAL fit in (c).

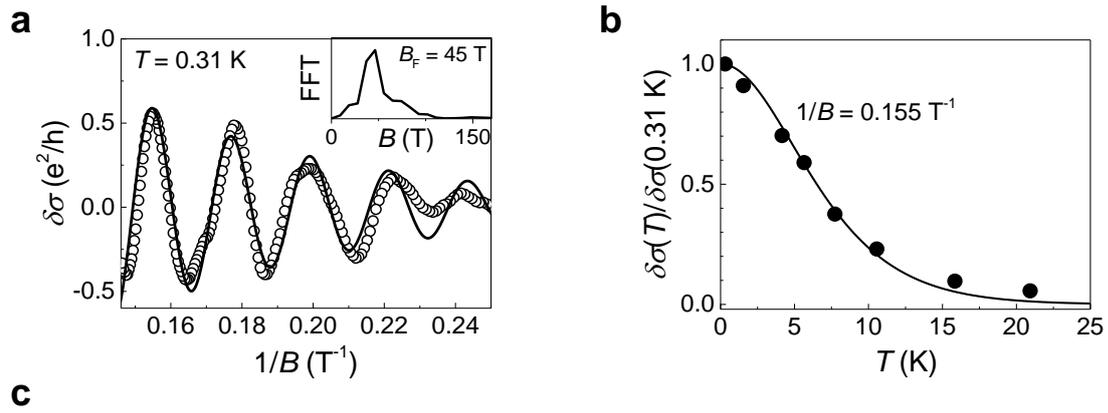

| $B_F$ (T) | $k_F$ (Å$^{-1}$) | $n_s$ (cm$^{-2}$) | $v_F$ (m/s) | $\mu$ (cm$^2$·V$^{-1}$·s$^{-1}$) | $m^*$ ($m_e$) | $l_e$ (nm) | $E_F$ (meV) |
|---|---|---|---|---|---|---|---|
| 45.06 ± 0.61 | 0.037 | 1.1 × 10$^{12}$ | 3 × 10$^5$ | 2095 | 0.147 ± 0.006 | 51 | 71 |

**Supporting Figure 7.** Shubnikov-de Haas (SdH) oscillations. (a) MC difference after subtracting the smooth background, $\delta\sigma$, as a function of the inverse magnetic field, $1/B$, applied perpendicular to the substrate. MC was measured at $T$ = 0.31 K and $V_g$ = 0 V for device D3. Solid line is a fit to the LK theory (see Supporting Note 2). Inset: FFT spectrum of $\delta\sigma(1/B)$ curve. (b) Normalized amplitude of SdH oscillations as a function of temperature for $1/B$ = 0.155 T$^{-1}$. The solid line is a fit to the LK theory. (c) Physical parameters obtained from the SdH oscillation analysis (see Supporting Note 2).

# Supporting Notes

## Supporting Note 1: Weak antilocalization (WAL) effect

We observed angle-dependent negative magnetoconductivity (MC), $\Delta\sigma(B)$, in Sb-doped $Bi_2Se_3$ TI NR, as shown in Supporting Fig. 6a, which is attributed to the weak antilocalization (WAL) effect[1]. The MC variation, $\Delta\sigma(B) = \sigma(B) - \sigma(0)$, became more negative when the perpendicular component of the magnetic field increased, and merged into a single curve at low magnetic fields of $B\sin\theta$ (see Supporting Fig. 6b). Our observations indicate that the topological surface states are mainly responsible for the negative MC behavior. The perpendicular MC curves for $\theta = 90°$ were measured at different temperatures, as shown in Supporting Fig. 6c. The low-field $\Delta\sigma(B)$ curve was fitted to the two-dimensional (2D) WAL theory[2], described by the Hikami-Larkin-Nagaoka (HLN) equation. According to the HLN theory, the MC variation is given as follows:[2]

$$\Delta\sigma = \sigma(B) - \sigma(0) = -\frac{\alpha e^2}{2\pi^2 \hbar}\left[\ln\left(\frac{\hbar}{4eBL_\varphi^2}\right) - \psi\left(\frac{1}{2} + \frac{\hbar}{4eBL_\varphi^2}\right)\right] \quad (1)$$

where $\alpha$ is a prefactor, $\hbar$ the reduced Planck's constant, $e$ the electronic charge, $\psi$ the digamma function, and $L_\varphi$ the phase coherence length. The results of fitting, $L_\varphi$ and $\alpha$ as a function of temperature, are shown in Supporting Figure 6d, revealing $L_\varphi = 465$ nm at $T = 3.0$ K. The red line means a power-law dependence of $L_\varphi$ on temperature, which is given by $L_\varphi \sim T^{-0.45}$.

## Supporting Note 2: Shubnikov-de Haas (SdH) oscillations

Supporting Fig. 7a shows magnetic quantum oscillations, which are known as Shubnikov-de Haas (SdH) oscillations, obtained at a perpendicular magnetic field ($\theta = 90°$). The MC difference, $\delta\sigma(B)$, was obtained after subtracting the smooth background from the MC, $\sigma(B)$, curve. The FFT analysis of $\delta\sigma(1/B)$ curve, as shown in the inset, results in an oscillation frequency of $B_F = 45$ T, corresponding to the oscillation period $B_F^{-1} = 2\pi e/\hbar S_F$, where $S_F = \pi k_F^2$ is the cross-sectional area of the Fermi surface and $k_F$ is the Fermi wave vector[3]. Thus we obtain $k_F = (2eB_F/\hbar)^{1/2} = 0.037$ Å$^{-1}$ and two-dimensional carrier concentration of $n_s = k_F^2/4\pi = 1.1 \times 10^{12}$ cm$^{-2}$. The solid line is a fit to the Lifshitz-Kosevich (LK) theory[4], which is expressed by $\delta\sigma = A\exp[-\pi/\mu B]\cos[2\pi(B_F/B + 1/2 + \beta)]$, where $A$ is a temperature-dependent parameter, $\mu$ the carrier mobility, and $2\pi\beta$ the Berry phase. The phase shift is expected to be $\beta = 0$ for a conventional 2D electron gas and $\beta = 0.5$ for 2D Dirac electrons[5]. From the LK fit, we obtain $\beta = 0.49$, indicating the existence of topological surface states in TI NR, and the surface electron mobility $\mu = 2095$ cm$^2 \cdot$V$^{-1} \cdot$s$^{-1}$.

Temperature dependence of the SdH oscillation amplitude is shown in Supporting Figure 7b for $1/B = 0.155$ T$^{-1}$. Since the amplitude parameter $A$ is proportional to $\lambda(T)/\sinh\lambda(T)$, where $\lambda(T) = 2\pi^2 k_B T m_c/\hbar eB$ is the thermal factor and $m_c$ is the cyclotron effective mass[6], we obtain $m_c = (0.147 \pm 0.006)m_e$, where $m_e$ is the electron mass, from the LK fitting (solid line). This value is in reasonable agreement with the previously reported values of $m_c = (0.12 \sim 0.16)\,m_e$ in Bi$_2$Se$_3$ NRs and nanoflakes[7, 8, 9]. As a result, we obtain the Fermi velocity $v_F = \hbar k_F/m_c = 3 \times 10^5$ m/s and the Fermi level $E_F = \hbar v_F k_F = m_c v_F^2 = 71$ meV.

## Supporting References

1. Chen J, *et al.* Gate-voltage control of chemical potential and weak antilocalization in $Bi_2Se_3$. *Phys Rev Lett* **105**, 176602 (2010).

2. Hikami S, Larkin AI, Nagaoka Y. Spin-orbit interaction and magnetoresistance in the two dimensional random system. *Progress of Theoretical Physics* **63**, 707-710 (1980).

3. Kim J, *et al.* Quantum Electronic Transport of Topological Surface States in *β*-$Ag_2Se$ Nanowire. *ACS Nano* **10**, 3936-3943 (2016).

4. Isihara A, Smrčka L. Density and magnetic field dependences of the conductivity of two-dimensional electron systems. *J Phys C Solid State Phys* **19**, 6777-6789 (1986).

5. Zhang YB, Tan YW, Stormer HL, Kim P. Experimental observation of the quantum Hall effect and Berry's phase in graphene. *Nature* **438**, 201-204 (2005).

6. Qu DX, Hor YS, Xiong J, Cava RJ, Ong NP. Quantum Oscillations and Hall Anomaly of Surface States in the Topological Insulator $Bi_2Te_3$. *Science* **329**, 821-824 (2010).

7. Kong D, *et al.* Rapid surface oxidation as a source of surface degradation factor for $Bi_2Se_3$. *ACS Nano* **5**, 4698-4703 (2011).

8. Tang H, Liang D, Qiu RLJ, Gao XPA. Two-dimensional transport-induced linear magneto-resistance in topological insulator $Bi_2Se_3$ nanoribbons. *ACS Nano* **5**, 7510-7516 (2011).

9. Fang L, *et al.* Catalyst-free growth of millimeter-long topological insulator $Bi_2Se_3$ nanoribbons and the observation of the π-berry phase. *Nano Lett* **12**, 6164-6169 (2012).